\documentclass[12pt,a4paper,final,nofootinbib]{revtex4}

\usepackage{sidecap}
\usepackage{paralist}
\usepackage{ulem}
\usepackage{epsfig}
\usepackage{amsmath,amssymb,amsthm}
\usepackage{graphicx}
\usepackage{bm}
\usepackage{color,soul}

\DeclareMathAlphabet{\mathpzc}{OT1}{pzc}{m}{it}

\begin{document}

\renewcommand{\textfraction}{0.00}


\newcommand{\pT}{p_\perp}
\newcommand{\vAi}{{\cal A}_{i_1\cdots i_n}}
\newcommand{\vAim}{{\cal A}_{i_1\cdots i_{n-1}}}
\newcommand{\vAbi}{\bar{\cal A}^{i_1\cdots i_n}}
\newcommand{\vAbim}{\bar{\cal A}^{i_1\cdots i_{n-1}}}
\newcommand{\htS}{\hat{S}}
\newcommand{\htR}{\hat{R}}
\newcommand{\htB}{\hat{B}}
\newcommand{\htD}{\hat{D}}
\newcommand{\htV}{\hat{V}}
\newcommand{\cT}{{\cal T}}
\newcommand{\cM}{{\cal M}}
\newcommand{\cMs}{{\cal M}^*}
\newcommand{\vk}{\vec{\mathbf{k}}}
\newcommand{\bk}{\bm{k}}
\newcommand{\kt}{\bm{k}_\perp}
\newcommand{\kp}{k_\perp}
\newcommand{\km}{k_\mathrm{max}}
\newcommand{\vl}{\vec{\mathbf{l}}}
\newcommand{\bl}{\bm{l}}
\newcommand{\bK}{\bm{K}}
\newcommand{\bb}{\bm{b}}
\newcommand{\qm}{q_\mathrm{max}}
\newcommand{\vp}{\vec{\mathbf{p}}}
\newcommand{\bp}{\bm{p}}
\newcommand{\vq}{\vec{\mathbf{q}}}
\newcommand{\bq}{\bm{q}}
\newcommand{\qt}{\bm{q}_\perp}
\newcommand{\qp}{q_\perp}
\newcommand{\bQ}{\bm{Q}}
\newcommand{\vx}{\vec{\mathbf{x}}}
\newcommand{\bx}{\bm{x}}
\newcommand{\tr}{{{\rm Tr\,}}}
\newcommand{\sNN}{s_{\mathrm{NN}}}
\newcommand{\bc}{\textcolor{blue}}

\newcommand{\beq}{\begin{equation}}
\newcommand{\eeq}[1]{\label{#1} \end{equation}}
\newcommand{\ee}{\end{equation}}
\newcommand{\bea}{\begin{eqnarray}}
\newcommand{\eea}{\end{eqnarray}}
\newcommand{\beqar}{\begin{eqnarray}}
\newcommand{\eeqar}[1]{\label{#1}\end{eqnarray}}

\newcommand{\half}{{\textstyle\frac{1}{2}}}
\newcommand{\ben}{\begin{enumerate}}
\newcommand{\een}{\end{enumerate}}
\newcommand{\bit}{\begin{itemize}}
\newcommand{\eit}{\end{itemize}}
\newcommand{\ec}{\end{center}}
\newcommand{\bra}[1]{\langle {#1}|}
\newcommand{\ket}[1]{|{#1}\rangle}
\newcommand{\norm}[2]{\langle{#1}|{#2}\rangle}
\newcommand{\brac}[3]{\langle{#1}|{#2}|{#3}\rangle}
\newcommand{\hilb}{{\cal H}}
\newcommand{\pleft}{\stackrel{\leftarrow}{\partial}}
\newcommand{\pright}{\stackrel{\rightarrow}{\partial}}

\title{Importance of higher harmonics and $v_4$ puzzle in quark-gluon plasma tomography}

\author{Dusan Zigic}
\affiliation{Institute of Physics Belgrade, University of Belgrade, Serbia}

\author{Jussi Auvinen}
\affiliation{Institute of Physics Belgrade, University of Belgrade, Serbia}

\author{Igor Salom}
\affiliation{Institute of Physics Belgrade, University of Belgrade, Serbia}

\author{Pasi Huovinen}
\affiliation{Institute of Physics Belgrade, University of Belgrade, Serbia}
\affiliation{Incubator of Scientific Excellence---Centre for Simulations of
             Superdense Fluids, University of Wroc\l{}aw, Poland}

\author{Magdalena Djordjevic\footnote{E-mail: magda@ipb.ac.rs}}
\affiliation{Institute of Physics Belgrade, University of Belgrade, Serbia}

\begin{abstract}
  QGP tomography aims to constrain the parameters characterizing the
  properties and evolution of Quark-Gluon Plasma (QGP) formed in
  heavy-ion collisions, by exploiting low and high-$p_\perp$ theory
  and data. Higher-order harmonics $v_n$ ($n>2$) are an
  important---but seldom explored---part of this approach. However, to
  take full advantage of them, several issues have to be addressed: i)
  consistency of different methods for calculating $v_n$, ii)
  importance of event-by-event fluctuations to high-$p_\perp$ $R_{AA}$
  and $v_2$ predictions, iii) sensitivity of higher harmonics to the
  initial state of fluid-dynamical evolution. We obtain that i)
  several methods for calculating harmonics are compatible with each
  other, ii) event-by-event calculations are important in mid-central
  collisions, and iii) various initializations of the evolution of the
  medium lead to quantitatively and qualitatively different
  predictions, likely to be distinguished by future measurements. We
  also find that the present high-$\pT$ $v_4$ data cannot be
  reproduced using initial states for fluid-dynamical evolution given
  by state-of-the-art models. We call this discrepancy high-$p_\perp$
  $v_4$ puzzle at the LHC.
\end{abstract}

\maketitle

\section{Introduction}

During the past two decades, an impressive experimental and theoretical effort has been invested in generating and exploring a new form of matter called Quark-Gluon Plasma (QGP)~\cite{QGP1,QGP2,QGP3,QGP4}. This form of matter consists of interacting and no longer confined quarks, antiquarks, and gluons~\cite{Collins,Baym} and is created at extremely high energy densities achieved in ultra-relativistic heavy ion collisions at the Relativistic Heavy Ion Collider (RHIC) and the Large Hadron Collider
(LHC) experiments. An unprecedented amount of data for different collision systems (large and small), collision energies, types of particles, momentum regions, centralities, etc., are generated in these experiments, and one of the major current goals is to optimally use these data to investigate the properties of this exciting form of matter.

As one of the latest experimental achievements, the high momentum
(high-$p_\perp$) higher harmonics have recently become available at
RHIC and the LHC. For example, for charged hadrons, the data are
available up to the $7^{th}$ harmonic (for ATLAS~\cite{ATLAS_CH_vn})
and cover the $p_\perp$ region up to 100 GeV (for CMS~\cite{CMS_CH_vn}).
For heavy flavor, the coverage is not that extensive (for both
harmonics and momentum region)---still, the upcoming experimental data
at high-luminosity LHC Run3 should provide these data for both light
and heavy flavor with much higher precision. In the upcoming RHIC
(sPHENIX and STAR) experiments, similar quality data is expected, with
$p_\perp$ coverage up to 20~GeV. Even if the $p_\perp$ range
accessible at RHIC is narrower than at the LHC, it is particularly
useful for QGP tomography due to the pronounced difference between
light and heavy flavor in that region. While these data (will) represent
the state-of-the-art in the experimental sector, theoretically the
higher harmonics at high-$\pT$ have not been well explored.

To use these data for QGP tomography, i.e., for exploring the bulk QGP
properties through high-$\pT$ theory and data, one should first
identify and address potential limitations, in particular related to
coverage and design of different experiments. For example, four
different methods are commonly used in the literature to evaluate
$v_n$: two-particle cumulant $v_n\{2\}$, four-particle cumulant
$v_n\{4\}$, event plane $v_n\{\mathrm{EP}\}$, and scalar product
$v_n\{\mathrm{SP}\}$ methods (see section~\ref{sec:analysis} for more
details). Do these methods provide consistent results, especially when
different experimental collaborations even define $v_n\{\mathrm{SP}\}$
in different ways?

Furthermore, in experimental analysis, the scalar product method
correlates the particle of interest at midrapidity with the bulk
medium constituents at higher rapidity regions to avoid non-flow
effects on measured $v_n$~\cite{ATLAS_CH_vn,CMS_CH_vn}. From
theoretical perspective, this means the use of the experimental
definition for $v_n\{\mathrm{SP}\}$ necessitates 3+1D hydrodynamic
modeling for event-by-event simulations. However, 3+1D simulations are
computationally several orders of magnitude more demanding than 2+1D
simulations and consequently time-wise impractical for high precision
QGP tomography. Thus, the question arises whether it would be
plausible to compare $v_n\{\mathrm{SP}\}$ obtained in boost-invariant
2+1D simulations to experimental data in a model where the high- and
low-$p_\perp$ particles have separate sources (fragmenting jets and a
thermal fireball, respectively), and are thus uncorrelated.

Next, for the second harmonic, $v_2$, event-by-event fluctuations are
expected to either have a significant effect on $v_2$
values~\cite{NoronhaHostler_ebe}, or to be small enough to be
considered negligible~\cite{He_ebe,CIBJET}. However, these studies
were done in limited and different centrality regions. It is
expected~\cite{Betz_ebe} that the effects of event-by-event
fluctuations increase with decreasing centrality. Thus, it is
important to systematically investigate and quantify these effects for
the high-$p_\perp$ region at different centralities.

Therefore, the study presented in this manuscript has the following main goals:
\begin{enumerate}[(i)]
\item Explore to what extent the different methods for calculating
  higher harmonics are compatible with each other.
\item Explore the importance of event-by-event fluctuations and
  correlations to high-$\pT$ $v_2$ and $R_{AA}$.
\item Explore the qualitative and quantitative effects of different
  medium evolution scenarios on high-$p_\perp$ higher harmonics, and
  how well the existing high-$\pT$ data can be reproduced without
  further tuning of parameters.
\end{enumerate}
Overall, this study explores whether and how high-$p_\perp$ higher
harmonics, with an adequate theoretical framework, can provide further
constraints to the bulk QGP properties.

\section{Methods}

  \subsection{Outline of DREENA-A framework}

To use the high-$\pT$ particles to explore the bulk properties, we
developed a fully optimized modular framework DREENA-A~\cite{DREENAA},
where ``DREENA'' stands for Dynamical Radiative and Elastic ENergy
loss Approach, while ``A'' stands for Adaptive. We further optimized
the framework for this study to efficiently incorporate any,
arbitrary, event-by-event fluctuating temperature profile within the
dynamical energy loss formalism. Due to the very large amount of
temperature profile data processed in event-by-event calculations,
we optimized file handling and formats. Also, we reorganized the
parallelization of computation, as well as ensured that
spatio-temporal resolution and calculation precision are optimal
and adjusted to the event-by-event type of profiles.

The framework does not
have fitting parameters within the energy loss model (i.e., all
parameters used in the model correspond to standard literature
values), which allows to systematically compare the data and
predictions obtained by the same formalism and parameter set.
Therefore different temperature profiles (which are the only input in
the DREENA-A framework) resulting from different initial states, and
QGP properties, can be distinguished by the high-$p_\perp$ observables
they lead to, and the bulk QGP properties can be further constrained
by studying low and high-$\pT$ theory and data jointly.

The dynamical energy loss formalism~\cite{MD_PRC,DH_PRL,MD_Coll} has several
important features, all of which are needed for accurate
predictions~\cite{BD_JPG}: {\it i)} QCD medium of {\it finite} size and
temperature consisting of dynamical (i.e., moving) partons.
{\it ii)} Calculations are based on generalized Hard-Thermal-Loop
approach~\cite{Kapusta}, with naturally regulated infrared
divergences~\cite{MD_PRC,MD_Coll,DG_TM}.
{\it iii)} Both radiative~\cite{MD_PRC,DH_PRL} and
collisional~\cite{MD_Coll} energy losses are calculated in the same
theoretical framework and apply to both light and heavy flavors.
{\it iv)} The framework is generalized toward running
coupling~\cite{MD_PLB} and finite magnetic mass~\cite{MD_MagnMass}.
We have also investigated the validity of the widely used
soft-gluon approximation~\cite{sga}, but found it a very good
approximation which does not need to be relaxed.

The initial quark spectrum, for light and heavy partons, is computed at next to leading order~\cite{Vitev0912}. We use DSS~\cite{DSS} fragmentation functions to generate charged hadrons, and BCFY~\cite{BCFY} and KLP~\cite{KLP} fragmentation
functions for D and B mesons, respectively. To generate high-$p_\perp$ predictions, we use the same parameter set
as in DREENA-A~\cite{DREENAA}. Specifically, we assume effective light quark flavors $n_f{\,=\,}3$ and $\Lambda_{QCD}=0.2$~GeV.
The temperature-dependent Debye mass
$\mu_E$ is obtained by applying the procedure from~\cite{Peshier} and
leads to results compatible with the lattice QCD~\cite{LatticeMass}.
For the gluon mass, we assume $m_g=\mu_E/\sqrt{2}$~\cite{DG_TM}, and for light quark mass $M{\,=\,}\mu_E/\sqrt{6}$.
The charm mass is $M{\,=\,}1.2$\,GeV and the bottom mass is $M{\,=\,}4.75$\,GeV. For magnetic to electric mass ratio, we use $\mu_M/\mu_E = 0.5$~\cite{Maezawa,Nakamura}.

  \subsection{Modeling the bulk evolution}
    \label{sec:BulkEvolution}

We investigate three different event-by-event initializations for the bulk evolution. The first is Monte Carlo Glauber (MC-Glauber) initialization at initial time $\tau_0 = 1.0$ fm without initial transverse flow. We assign the binary collision points at halfway between the two colliding nucleons and convert these points to a continuous binary collision density using 2-D Gaussian distributions
\begin{equation}
  n_{BC}(x,y) = \frac{1}{2\pi\sigma_{BC}^2}
                \sum_{i=1}^{N_{BC}}\exp\left(-\frac{(x-x_i)^2+(y-y_i)^2}
                                                {2\sigma_{BC}^2} \right)
\end{equation}
with a width parameter $\sigma_{BC}=0.35$ fm. The binary collision density is then converted to energy density with the formula
\begin{equation}
 \epsilon(x,y)=C_0(n_{BC} + c_1 n_{BC}^2 + c_2 n_{BC}^3),\label{energydensity}
\end{equation}
and further extended in the longitudinal direction using the LHC parametrization from Ref.~\cite{Molnar:2014zha}.
The evolution of the fluid is calculated using a 3+1D viscous fluid code~\cite{Molnar:2014zha}, with a constant shear viscosity over entropy density ratio $\eta/s = 0.03$ and no bulk viscosity. The equation of state (EoS) parametrization is
$s95p$-PCE-v1~\cite{Huovinen:2009yb}. The model parameters were tuned to ALICE charged particle multiplicity \cite{Adam:2015ptt} and $v_n(p_\perp)$ data \cite{Adam:2016izf} for 10-20\%, 20-30\% and 30-40\% centrality classes in Pb+Pb collisions at $\sqrt{s_{NN}}=5.02$ TeV.

The second model is the T$_\mathrm{R}$ENTo
initialization~\cite{Moreland:2014oya} with a free streaming stage until $\tau_0 = 1.16$ fm, further evolved using the VISH2+1
code~\cite{Song:2007ux} as described in~\cite{Bernhard:2018hnz,
  Bernhard:2019bmu}. The parameters in this calculation are
based on a Bayesian analysis of the data at Pb+Pb collisions at $\sqrt{s_{NN}}=2.76$ and $5.02$ TeV~\cite{Bernhard:2019bmu}. In particular the calculation
includes temperature dependent shear and bulk viscosity coefficients with the minimum value of $\eta/s = 0.081$ and maximum of $\zeta/s = 0.052$. The EoS \cite{Bernhard:2018hnz} is based on
the lattice results by the HotQCD collaboration~\cite{HotQCD:2014kol}.

The third investigated initialization model is IP-Glasma~\cite{ipglasma,ipglasma2}. The calculated event-by-event fluctuating initial states~\cite{Schenke:2020mbo} are further evolved
~\cite{Chun-private}
using the MUSIC code~\cite{Schenke:2010nt,Schenke:2010rr,Schenke:2011bn} constrained to boost-invariant expansion.
In these calculations, the switch from Yang-Mills to fluid-dynamical evolution takes place at $\tau_{\mathrm{switch}}=0.4$ fm, shear viscosity over entropy density ratio is constant $\eta/s=0.12$, and the temperature-dependent bulk viscosity coefficient over entropy density ratio has the maximum value $\zeta/s = 0.13$. The equation of state is based on the HotQCD lattice results~\cite{HotQCD:2014kol} as presented in Ref.~\cite{Moreland:2015dvc}.

   \subsection{Flow analysis}
        \label{sec:analysis}

      \subsubsection{Scalar product and event plane methods}

We start by defining the low-$p_\perp$ normalized flow vector for
$n$-th harmonic based on $M$ particles as
\begin{equation}
 Q_n=\frac{1}{M} \sum_{j=1}^{M} e^{in\phi_j} \equiv |v_n| e^{in\Psi_n},
\end{equation}
where $\Psi_n$ is the event plane angle:
$\Psi_n=\arctan(\frac{\mathrm{Im}\,Q_n}{\mathrm{Re}\,Q_n}) / n$.

Similarly to low $p_\perp$, we can define the flow vector for a high
$p_\perp$ bin as ($R_{AA}(p_\perp) =\frac{1}{2\pi} \int_0^{2\pi} R_{AA} (p_\perp,\phi)\,d\phi $)
\begin{equation}
 q_n^{\mathrm{hard}}=\frac{\frac{1}{2\pi}\int_0^{2\pi}\, e^{in\phi} R_{AA}(p_\perp,\phi)\,d\phi}{R_{AA}(p_\perp)},
\end{equation}
and single-event high-$p_\perp$ flow coefficients $v_n^{\mathrm{hard}}$ as \cite{NoronhaHostler_ebe}
\begin{equation}
v_n^{\mathrm{hard}}=\frac{\frac{1}{2\pi}\int_0^{2\pi}\,\cos[n(\phi-\Psi_n^{\mathrm{hard}}(p_\perp))] R_{AA}(p_\perp,\phi)\,d\phi}{R_{AA}(p_\perp)}
\end{equation}
where the event plane angle $\Psi_n^{\mathrm{hard}}(p_\perp)$ is defined as
\begin{equation}
 \Psi_n^{\mathrm{hard}}(p_\perp) = \frac{1}{n}\arctan\left( \frac{\int_0^{2\pi}\,\sin(n\phi)R_{AA}(p_\perp,\phi)\,d\phi}{\int_0^{2\pi}\,\cos(n\phi)R_{AA}(p_\perp,\phi)\,d\phi} \right).
\end{equation}
The high-$p_\perp$ $v_n$ is then calculated by correlating $q_n$ with $Q_n$ \cite{NoronhaHostler_ebe,Andres:2019eus,He_ebe}:
\begin{equation}
v_n^{\mathrm{hard}}\{\mathrm{SP}\}=\frac{\langle \mathrm{Re}\,(q_n^{\mathrm{hard}}(Q_n)^*) \rangle_{\mathrm{ev}}}{\sqrt{\langle Q_n (Q_n)^* \rangle_{\mathrm{ev}}}}\\
=\frac{\langle |v_n^{\mathrm{hard}}| |v_n|\cos[n(\Psi_n^{\mathrm{hard}}(p_\perp)-\Psi_n)] \rangle_{\mathrm{ev}}}{\sqrt{\langle |v_n|^2 \rangle_{\mathrm{ev}}}}.
\label{eq:SPmidr}
\end{equation}
We may also simply calculate the high-$p_\perp$ anisotropy with respect to the event plane $\Psi_n$, which we shall denote as the ``event plane'' $v_n$ \cite{He_ebe}:
\begin{equation}
 v_n\{\mathrm{EP}\}=\langle \langle \cos[n(\phi^{\mathrm{hard}}-\Psi_n)] \rangle \rangle_{ev} = \langle v_n^{\mathrm{hard}}\cos[n(\Psi^{\mathrm{hard}}_n-\Psi_n)] \rangle_{ev}.
 \label{eq:EP}
\end{equation}

For our theoretical $v_n\{\mathrm{SP}\}$, the reference flow vector
$Q_n$ is calculated using only midrapidity particles. In order to
reduce non-flow effects, it is common in experiments to introduce a
rapidity gap between the particles of interest and the reference flow
particles. ATLAS defines the scalar product $v_n$ as~\cite{ATLAS_CH_vn}:
\begin{equation}
 v_n\{\mathrm{SP}_{\mathrm{ATLAS}}\}=\frac{\mathrm{Re}\, \langle \langle e^{in\phi} (Q_n^{-|+})^* \rangle \rangle_{\mathrm{ev}}}{\sqrt{\langle Q_n^- (Q_n^+)^* \rangle}_{\mathrm{ev}}}
 \label{eq:SPATLAS}
\end{equation}
where $Q_n^-=\frac{1}{M^-}\sum_{j=1}^{M^-} e^{in\phi_j}$ refers to particles in the rapidity interval $-4.9 < \eta < -3.2$ and $Q_n^+$ similarly to particles in the interval $3.2 < \eta < 4.9$, while $e^{in\phi}$ is associated with particles in midrapidity $|\eta|<2.5$. $Q_n^{-|+}$ indicates that particle of interest with $\eta<0$ are coupled to $Q_n^+$ and particles with $\eta>0$ to $Q_n^-$ to maximize the rapidity gap~\footnote{Since our high-$p_\perp$ particles are produced at $\eta=0$, the choice of $Q_n^+$ or $Q_n^-$ for the correlation is arbitrary.}.

CMS definition for the scalar product is \cite{CMS_CH_vn}
\begin{equation}
 v_n\{\mathrm{SP}_{\mathrm{CMS}}\}=\frac{\mathrm{Re}\, \langle Q_n Q_{nA}^* \rangle_{\mathrm{ev}}}{\sqrt{\frac{
 \langle Q_{nA} Q_{nB}^* \rangle_{\mathrm{ev}}
 \langle Q_{nA} Q_{nC}^* \rangle_{\mathrm{ev}}}
 {\langle Q_{nB} Q_{nC}^* \rangle_{\mathrm{ev}}}}}
 \label{eq:SPCMS}
\end{equation}
where the flow vector $Q_n=\sum_{j=1}^{M} e^{in\phi_j}$ consists of
particles of interest in midrapidity $|\eta|<1.0$, vectors
$Q_{nA},Q_{nB}=\sum_{j=1}^{M_{A,B}} E_{T}\,e^{in\phi_j}$ are measured from
the HF calorimeters at $2.9 < |\eta| < 5.2$, one at the negative and the
other at the positive rapidity, and the third reference vector
$Q_{nC}=\sum_{j=1}^{M_C} p_\perp\,e^{in\phi_j}$ is obtained from tracks with
$|\eta|<0.75$. If the particle of interest comes from the positive-$\eta$
side of the tracker, then $Q_{nA}$ is calculated using the negative-$\eta$
side of HF, and vice versa.

      \subsubsection{Cumulant method}

For 2- and 4-particle cumulant analysis, we use the unnormalized flow vector:
\begin{equation}
 \tilde{Q}_n=\sum_{j=1}^{M} e^{in\phi_j}.
\end{equation}

The low-$p_\perp$ integrated reference flow is calculated using
Eqs.~(7)-(18) from Ref.~\cite{Bilandzic:2010jr}: The 2-particle
cumulant $v_n$ is defined as
\begin{equation}
 v_n\{2\}=\sqrt{c_n\{2\}},
 \label{eq:2pCumulant}
\end{equation}
where the second order cumulant $c_n\{2\}$ equals the event-averaged 2-particle correlation $\langle \langle 2 \rangle \rangle_{\mathrm{ev}}$. The 4-particle cumulant $v_n$ is
\begin{equation}
 v_n\{4\}=\sqrt[4]{-c_n\{4\}},
 \label{eq:4pCumulant}
\end{equation}
where $c_n\{4\}$ is the 4th order cumulant $\langle \langle 4 \rangle \rangle_{\mathrm{ev}}-2\langle\langle 2 \rangle \rangle^2_{\mathrm{ev}}$.

For a single event, the 2-particle correlation is
\begin{equation}
 \langle 2 \rangle = \frac{|\tilde{Q}_n|^2-M}{W_2}
\end{equation}
with a combinatorial weight factor $W_2=M(M-1)$ and the single-event 4-particle correlation is
\begin{equation}
\begin{split}
 \langle 4 \rangle = &\frac{|\tilde{Q}_n|^4+|\tilde{Q}_{2n}|^2-2\mathrm{Re}|\tilde{Q}_{2n}\tilde{Q_n^*}\tilde{Q_n^*}|}{W_4}\\
 &-2\frac{2(M-2)|\tilde{Q}_n|^2-M(M-3)}{W_4}
 \end{split}
\end{equation}
with $W_4=M(M-1)(M-2)(M-3)$.

Using the weight factors defined above, the weighted average of a $k$-particle correlation over multiple events is then
\begin{equation}
 \langle \langle k \rangle \rangle_{\mathrm{ev}} = \frac{\sum\limits_{i=1}^{N_\mathrm{events}} W_{k,i} \,  \langle k \rangle_i}{\sum\limits_{i=1}^{N_\mathrm{events}} W_{k,i}}.
\end{equation}

Once the reference flow has been determined, the $p_T$-differential flow can be calculated using Eqs.~(20)-(35) of~\cite{Bilandzic:2010jr}. Here we denote the flow vector in a $p_\perp$ bin with $m_q$ particles as
\begin{equation}
 q_n=\sum_{j=1}^{m_q} e^{in\phi_j}.
\end{equation}
For high-$p_\perp$ particles, $q_n$ is calculated from the distribution
\begin{equation}
\label{eq:highptqn}
 q_n=\int_{0}^{2\pi} e^{in\phi} \frac{dN}{dp_\perp d\phi} d\phi
\end{equation}
with the associated multiplicity
\begin{equation}
\label{eq:highptmq}
 m_q=\int_{0}^{2\pi} \frac{dN}{dp_\perp d\phi} d\phi.
\end{equation}
For high-$p_\perp$ differential flow, none of the particles in a
$p_\perp$ bin are included in the calculation of the reference flow,
so the weight factors are $W'_2=m_qM$ and $W'_4=m_qM(M-1)(M-2)$, and
the 2-particle correlation is simply
\begin{equation}
\label{eq:highpttwocorr}
\langle 2' \rangle = \frac{q_n\tilde{Q}_n^*}{W'_2},
\end{equation}
while the 4-particle correlation is
\begin{equation}
  \langle 4' \rangle = \frac{q_n \tilde{Q}_n \tilde{Q}_n^* \tilde{Q}_n^* -q_n \tilde{Q}_n \tilde{Q}_{2n}^* -2Mq_n\tilde{Q}_n^* +2q_n\tilde{Q}_n^*}{W'_4}.
\end{equation}
With the knowledge of the correlations, we can calculate the differential cumulants
\begin{equation}
 \begin{split}
 d_n\{2\}&=\langle\langle 2' \rangle \rangle_{\mathrm{ev}}, \\
 d_n\{4\}&=\langle\langle 4' \rangle \rangle_{\mathrm{ev}} - 2\langle\langle 2' \rangle \rangle_{\mathrm{ev}} \langle\langle 2 \rangle \rangle_{\mathrm{ev}}
 \end{split}
\end{equation}
and the differential flow
\begin{equation}
 \begin{split}
  v'_n\{2\}=\frac{d_n\{2\}}{\sqrt{c_n\{2\}}}, \\
  v'_n\{4\}=-\frac{d_n\{4\}}{(-c_n\{4\})^{3/4}}.
 \end{split}
\end{equation}

\section{Results and discussion}

   \subsection{Compatibility of analysis methods}

\begin{figure*}
\centering
\epsfig{file=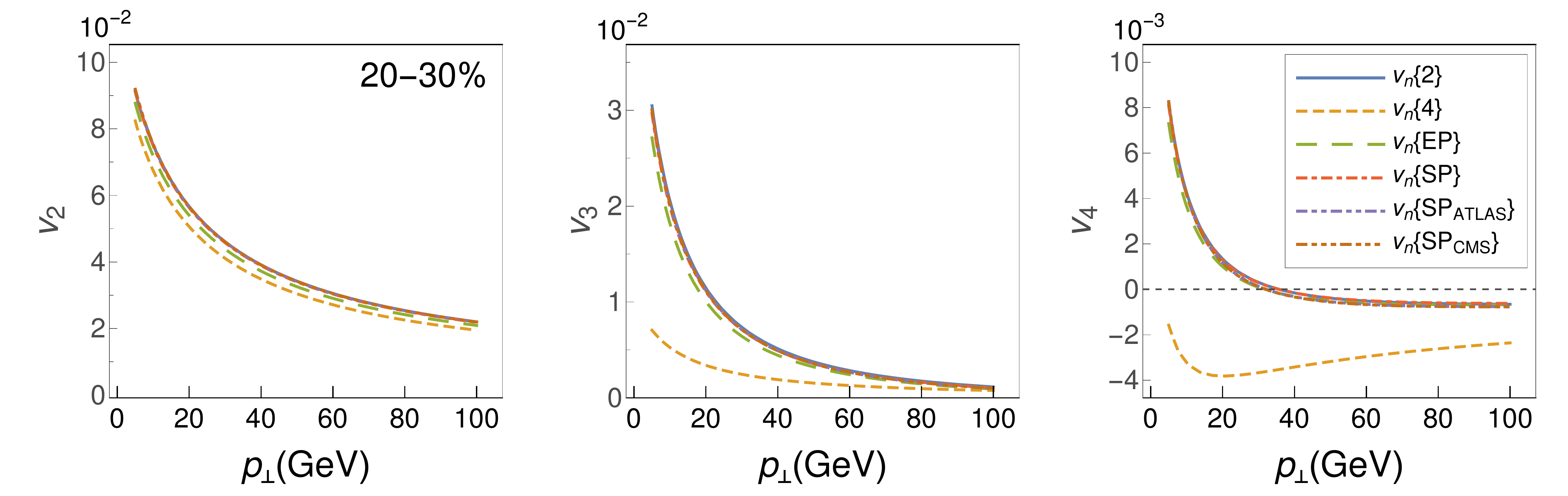,scale=0.35}
\vspace*{-0.2cm}
 \caption{Charged hadron $v_{2}$ (left), $v_3$ (middle) and $v_4$
   (right) in Pb+Pb collisions at $\sqrt{\sNN} = 5.02$~TeV for
   20-30\% centrality class, computed using different analysis
   methods: 2-particle cumulant, 4-particle cumulant, event plane,
   midrapidity scalar product, ATLAS-defined scalar product, and CMS
   defined scalar product, each described in the section~\ref{sec:analysis}. Energy
   loss calculation was performed on MC-Glauber+3d-hydro temperature profiles,
   with $\mu_M/\mu_E=0.5$.}
 \label{fig:VnAvgMethods}
\end{figure*}

In Fig.~\ref{fig:VnAvgMethods}, we compare $v_n(p_\perp)$ for
high-$p_\perp$ particles obtained using six different methods:
2-particle cumulant $v_n\{2\}$ given by Eq.~(\ref{eq:2pCumulant}),
4-particle cumulant $v_n\{4\}$ given by Eq.~(\ref{eq:4pCumulant}),
event plane $v_n\{\mathrm{EP}\}$ defined by Eq.~(\ref{eq:EP}),
midrapidity scalar product $v_n\{\mathrm{SP}\}$ calculated using
Eq.~(\ref{eq:SPmidr}), scalar product $v_n\{\mathrm{SP}_{\mathrm{ATLAS}}\}$
as defined by the ATLAS collaboration (Eq.~(\ref{eq:SPATLAS})), and scalar
product $v_n\{\mathrm{SP}_{\mathrm{CMS}}\}$ as defined by the CMS
collaboration (Eq.~(\ref{eq:SPCMS})). High-$p_\perp$ $R_{AA}$ and
$v_n$ predictions were obtained using generalized DREENA-A
framework with the temperature profiles calculated using the combination
of 3+1D viscous fluid code and MC-Glauber initial conditions
(i.e., the first bulk model described in the section~\ref{sec:BulkEvolution}).

As illustrated in Fig.~\ref{fig:VnAvgMethods}, different scalar
product methods for evaluating the $v_n$ coefficients, and the
2-particle cumulant method, lead to the same results with $\approx 5$\% level
accuracy. In agreement with Refs.~\cite{He_ebe,Shen_Photons,CIBJET},
the event plane results are also comparable to the scalar product
results deviating only $\approx 10$\%, i.e., less than the current
experimental uncertainty. The only method with significantly different
results is the four-particle cumulant method $v_n\{4\}$, which is
expected to differ from $v_n\{2\}$ in the presence of event-by-event
fluctuations~\cite{Betz_ebe,Voloshin:2007pc}. The equivalence of
different approaches simplifies comparison between theoretical
predictions and experimental results, since a theoretical prediction
calculated using any method (with the exception of the 4-particle
cumulant method) can be directly compared to experimental data
analyzed using any method. We have also checked that, in the scalar
product method, the rapidity of particles used to calculate the
reference flow vector has a negligible impact on high-$p_{\perp}$
particle $v_n$ in our framework and setup, allowing us to make
meaningful $v_n\{\mathrm{SP}\}$ data comparisons using the
boost-invariant hydro simulations. However, it must be remembered that
the scalar product method with large rapidity gap can be affected by
the event plane decorrelation at different rapidities~\cite{Jia:2014vja,CMS:2015xmx}.
In our approach the event plane is the same independent of rapidity,
and thus the effect of decorrelation is not included. How the event
plane depends on rapidity depends on the model used to create the
longitudinal structure of the initial state, and since there are very
few theoretical constraints for it, we leave these studies for a later
work.

  \subsection{Event-by-event fluctuations}

\begin{figure*}
\centering
\epsfig{file=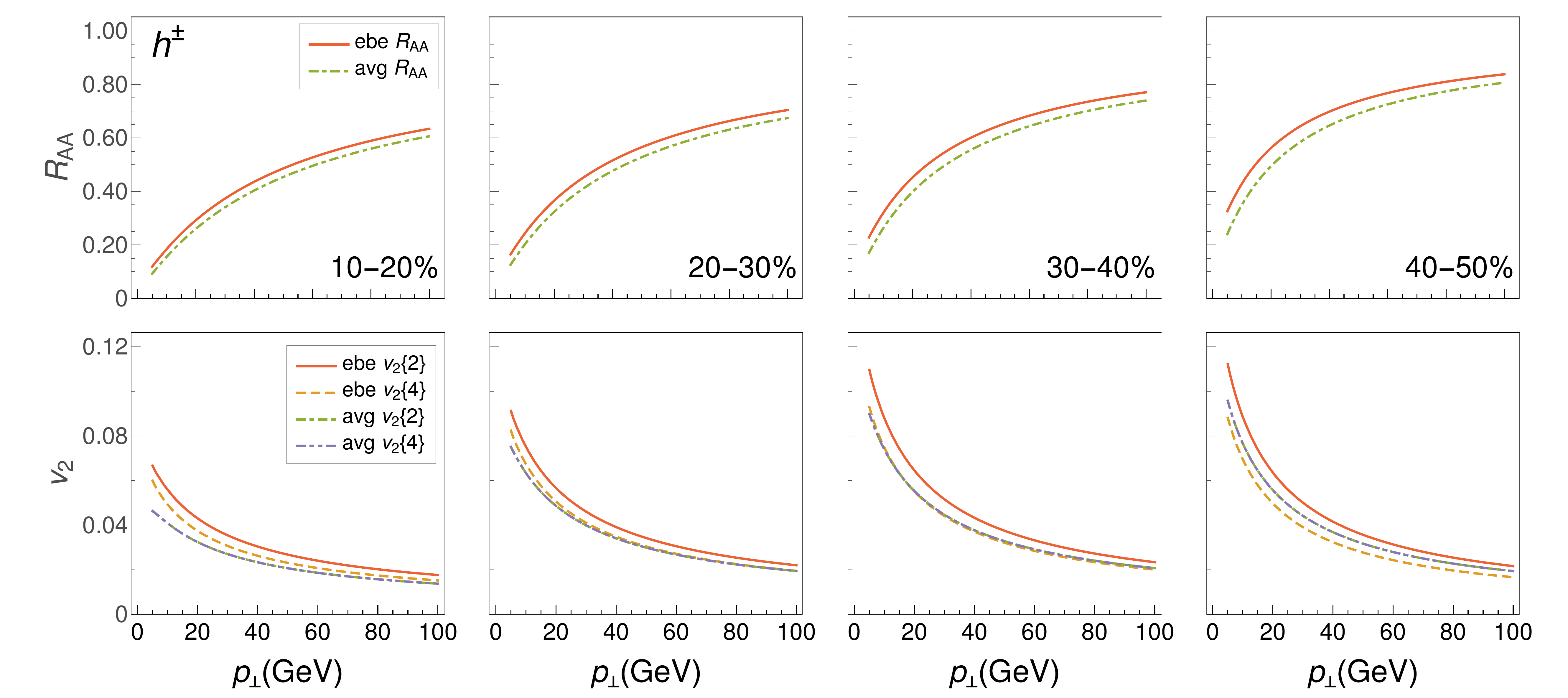,scale=0.30}
\vspace*{-0.2cm}
  \caption{ \textit{Upper panels:} charged hadron $R_{AA}$ calculated
    using event-by-event (ebe) fluctuating temperature profiles compared to
    $R_{AA}$ calculated using a  smooth temperature profile
    (avg). \textit{Lower panels:} charged hadron $v_n\{2\}$ and
    $v_n\{4\}$ calculated using event-by-event (ebe) fluctuating temperature
    profiles compared to $v_n\{2\}$ and $v_n\{4\}$ calculated using a
    smooth temperature profile (avg). Calculation was done for Pb+Pb
    collisions at $\sqrt{\sNN} = 5.02$~TeV, $\mu_M/\mu_E=0.5$, using
    MC-Glauber+3d-hydro bulk evolution.
    Each column represents different centrality class (from left to
    right: 10-20\%, 20-30\%, 30-40\% and 40-50\%).}
\label{fig:EbEvsAVG}
\end{figure*}

To investigate the influence of event-by-event fluctuations on
high-$p_\perp$ observables, MC-Glauber initial conditions for all
events within a single centrality class were averaged (we kept
reaction planes aligned, and averaged binary collision densities
before converting to energy density, (Eq.~\ref{energydensity})) and then
evolved using the 3+1D viscous fluid code (in a single run, instead of
one run for each event). Obtained smooth temperature profile was used to
calculate high-$p_\perp$ predictions, and $R_{AA}$ as well as
$v_2\{2\}$ and $v_2\{4\}$ results were compared to those obtained
using full event-by-event calculations (evolved separately for each
event), see Fig~\ref{fig:EbEvsAVG}.

We see that event-by-event fluctuations increase both $R_{AA}$ and $v_2$. While the effect on the $R_{AA}$ values is rather small ($\approx 7\%$) and does not have clear centrality dependence, the effect on $v_2\{2\}$ is more pronounced and increases with decreasing centrality. Quantitatively, we obtain that the average difference between event-by-event $v_2\{2\}$ and $v_2\{2\}$ calculated using smooth temperature profile goes from 14\% for the 40-50\% centrality class to 32\% in the 10-20\% centrality class.
The observed centrality dependence can be explained by the fact that with the increase in centrality, the influence of geometry on $v_2\{2\}$ becomes larger, while at low centralities, event-by-event fluctuations have the dominant impact on $v_2\{2\}$. We also observe a $p_\perp$ dependence of these differences (generally decreasing with increasing $p_\perp$) and no notable difference between $v_2\{2\}$ and $v_2\{4\}$ when calculated on the smooth temperature profile, where initial state eccentricity fluctuations are absent.

  \subsection{Effects of initial state}

\begin{figure*}
\centering
\epsfig{file=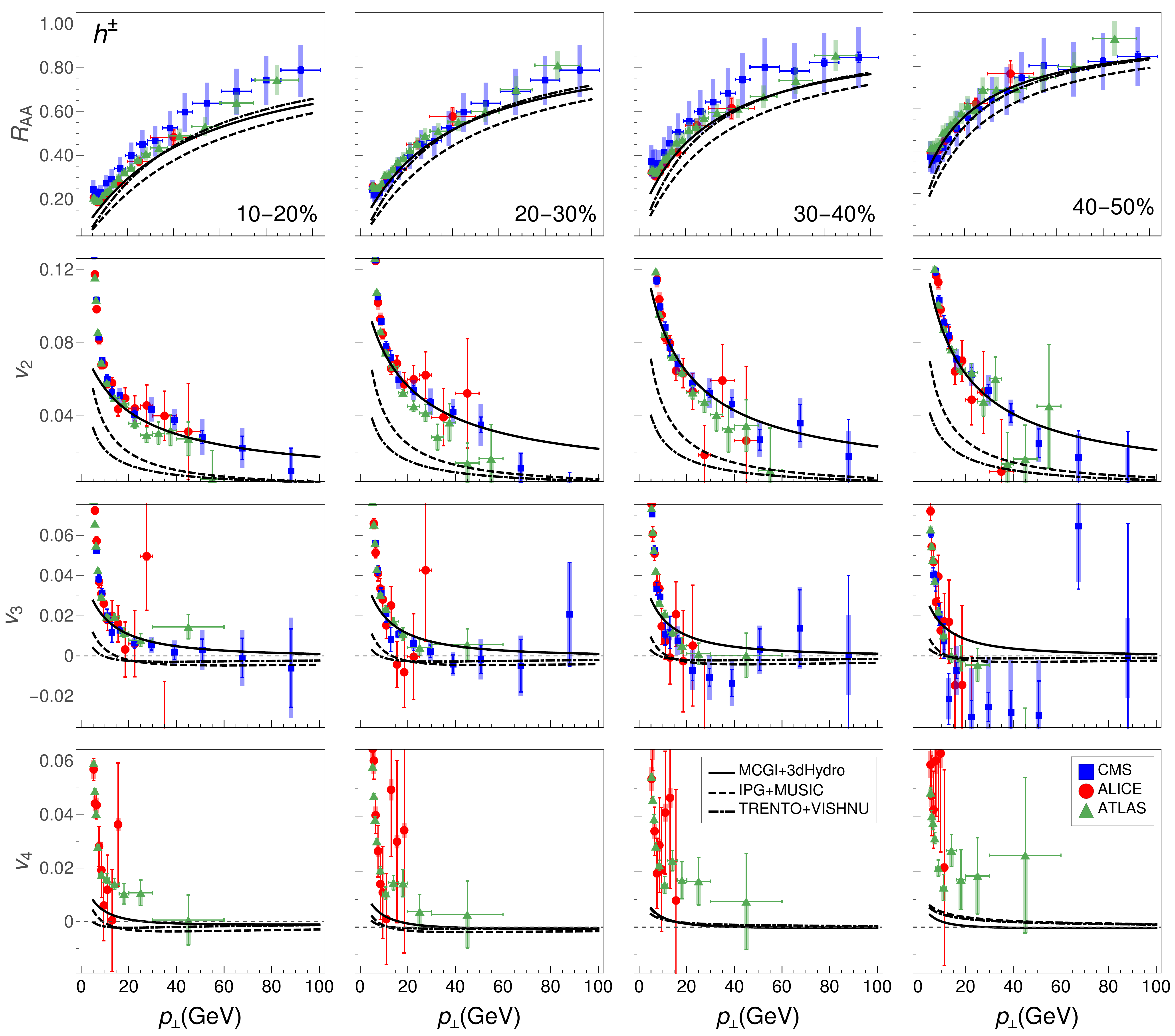,scale=0.30}
\vspace*{-0.2cm}
 \caption{Charged hadron $R_{AA}$ (first row) $v_{2}$ (second row),
   $v_3$ (third row) and $v_4$ (fourth row) in Pb+Pb collisions at
   $\sqrt{\sNN} = 5.02$~TeV for different initializations of the QGP evolution
   (indicated in the legend). Theoretical predictions,
   obtained using SP method, are compared to
   CMS~\cite{CMS_CH_RAA,CMS_CH_vn} (blue squares),
   ALICE~\cite{ALICE_CH_RAA,ALICE_CH_vn} (red circles) and
   ATLAS~\cite{ATLAS_CH_RAA,ATLAS_CH_vn} (green triangles)
   data. Columns 1-4 correspond to, respectively,
   10-20\%, 20-30\%, 30-40\% and 40-50\% centrality
   classes. $\mu_M/\mu_E=0.5$.}
 \label{fig:CHExpComp}
\end{figure*}

\begin{figure*}
\centering
\epsfig{file=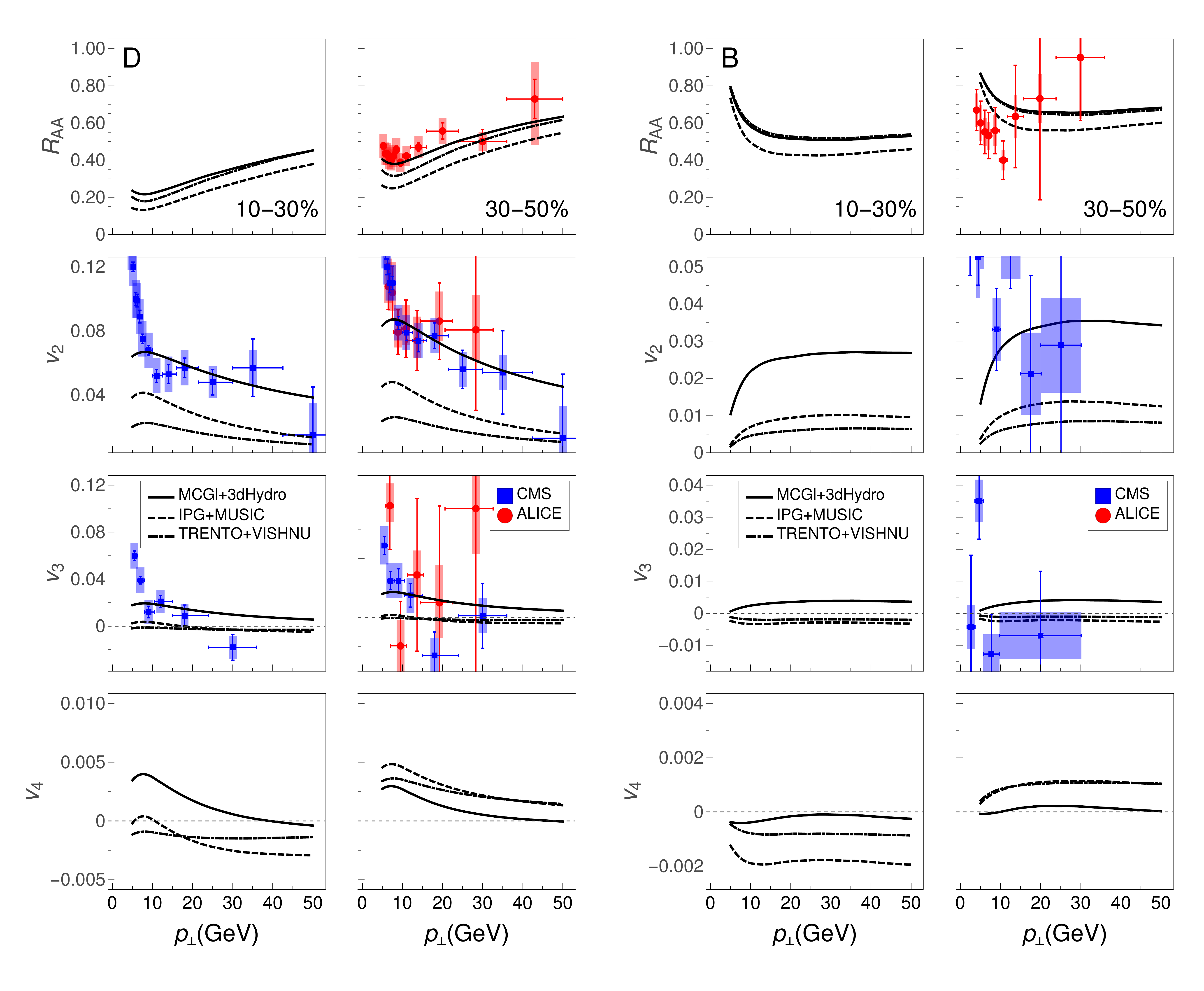,scale=0.30}
\vspace*{-0.2cm}
 \caption{D meson (left $4\times2$ panel) and B meson (right $4\times2$ panel) predictions in Pb+Pb collisions at $\sqrt{\sNN} = 5.02$~TeV for different initializations of QGP evolution (indicated in the legend). In each $4\times2$ panel, first row corresponds to $R_{AA}$, the second, third, fourth to $v_2$, $v_3$, $v_4$, respectively, while the left (right) column corresponds to 10-30\% (30-50\%) centrality class. D meson theoretical predictions are compared to CMS~\cite{CMS_D_vn} (blue squares) and ALICE~\cite{ALICE_D_RAA,ALICE_D_vn} (red circles) data, while B meson predictions are compared to preliminary CMS~\cite{CMS_B_vn} (blue squares) and preliminary ALICE~\cite{ALICE_B_RAA} (red circles) data for non-prompt D meson from b decay. $\mu_M/\mu_E=0.5$.
 }
\label{fig:DBExpComp}
\end{figure*}

To demonstrate the applicability of high-$p_\perp$ theoretical predictions as a QGP tomography tool, we generated three different sets of temperature profiles using three different initial conditions and hydrodynamics codes. Generalized DREENA-A~\cite{DREENAA} was then used to calculate high-$p_\perp$ predictions, which are compared to experimental data and, for charged hadrons, presented in Fig.~\ref{fig:CHExpComp}, and for D and B mesons in Fig.~\ref{fig:DBExpComp}. As can be seen, different initializations of fluid-dynamical evolution lead to different high-$p_\perp$ predictions for both $R_{AA}$ and $v_{2}$, $v_{3}$ and $v_{4}$, even though they all provide good agreement with low-$p_\perp$ data. Specific differences are visible already on the level of $R_{AA}$ values, where the IP-Glasma model results in discernibly stronger suppression. The differences in predictions become even higher when we consider the $v_2$ observable, with T$_\mathrm{R}$ENTo leading to lower $v_2$ than IP-Glasma, while MC-Glauber predictions are far above the two. A similar magnitude of relative differences is also obtained for $v_3$ and $v_4$ predictions, with an additional qualitative signature appearing for these observables: we notice that some initializations lead to negative values of high-$\pT$ $v_3$ and $v_4$, i.e., models can differ even in the expected sign of the flow coefficients.

Since DREENA-A does not have fitting parameters in the energy loss (the only inputs are the temperature profile and binary collisions, which come as a direct output from fluid-dynamical calculation and the initial state model), Figs.~\ref{fig:CHExpComp} and~\ref{fig:DBExpComp} demonstrate that high-$p_\perp$ $R_{AA}$ and higher harmonics can distinguish between different initializations and temperature profiles, and subsequently further constrain their parameters. Furthermore, Fig.~\ref{fig:DBExpComp} suggests that heavy flavor high-$p_\perp$ observables are even more sensitive to different temperature profiles than the light flavor. We also see that predictions for high-$p_\perp$ higher harmonics can be either positive or negative. Thus, the high-$p_\perp$ sector can provide both quantitative and qualitative constraints for different initial states.

Presently, of the considered models, the best agreement is observed
for MC-Glauber. This result is compatible with our earlier
findings~\cite{Stojku:2020wkh}, where the best agreement with
high-$p_\perp$ data was found by delaying the start of transverse
expansion and energy loss to time $\tau_0\approx 1.0$ fm. However, all models
seem to vastly underestimate the $v_4$ values, though the error bars for
the available $v_4$ data are quite large. If this tendency is
preserved in future high luminosity experiments (e.g., in LHC run 3),
it will present a new ``high-$p_\perp$ $v_4$ puzzle'', whose solution
will require modifications to the present initial state models and/or energy
loss mechanisms. Additionally, better quality heavy flavor data are
needed, especially D and B-meson data, as they present valuable
constraint to the evolution of the medium.

\section{Summary}

We obtained four main conclusions in this work: {\it i}) We found that
different methods to calculate higher harmonics at high-$p_\perp$ are
compatible with each other within $\approx 5$--10$\%$ accuracy, which is
less than the current experimental uncertainties.
%
%
{\it ii}) Event-by-event calculations are
particularly important for high-$p_\perp$ $v_2$ in mid-central
collisions. {\it iii}) Predictions for high-$p_\perp$ observables, and
especially for higher harmonics, are sensitive to the initial state of
fluid-dynamical evolution, and can distinguish between different
initial state models. {\it iv}) All initial state models lead to way
smaller high-$p_\perp$ $v_4$ than experimentally observed, and this
disparity deserves to be called a ``$v_4$ puzzle''. Overall, the
higher harmonics provide an exciting opportunity to obtain further
constraints to the QGP properties and its evolution in heavy-ion
collisions by combining new theoretical developments (with the
corresponding predictions) and upcoming higher luminosity experimental
measurements.

{\em Acknowledgments:}
We thank Chun Shen for sharing his results with us. We thank Marko
Djordjevic, Bojana Ilic and Stefan Stojku for useful discussions. This
work is supported by the European Research Council, grant
ERC-2016-COG: 725741, and by the Ministry of Science and Technological
Development of the Republic of Serbia. PH was also supported by the
program Excellence Initiative--Research University of the University of
Wroc\l{}aw of the Ministry of Education and Science.

\end{document}